# Crystal growth, transport phenomena and two gap superconductivity in the mixed alkali metal $(K_{1-z}Na_z)_xFe_{2-y}Se_2$ iron selenide

Maria Roslova[a], Svetoslav Kuzmichev[*,b], Tatiana Kuzmicheva[b,c], Yevgeny Ovchenkov[b], Min Liu[a,b], Igor Morozov[a], Aleksandr Boltalin[a], Andrey Shevelkov[a], Dmitry Chareev[d], Alexander Vasiliev[b,e]

[a] Department of Inorganic Chemistry, Faculty of Chemistry, Lomonosov Moscow State University, 119991 Moscow, Russia

[b] Low Temperature Physics and Superconductivity Department, Physics Faculty, M.V. Lomonosov Moscow State University, Moscow 119991, Russia

[c] P.N. Lebedev Physical Institute of the RAS, 119991 Moscow, Russia

[d] Institute of Experimental Mineralogy, Russian Academy of Sciences, 142432 Chernogolovka, Moscow District, Russia

[e] Theoretical Physics and Applied Mathematics Department, Institute of Physics and Technology, Ural Federal University, Ekaterinburg 620002, Russia

[*] Corresponding author E-mail: kuzmichev at mig.phys.msu.ru;
Fax: +7 (495) 9329217; Tel: +7 (495) 9329217

Abstract

Using the self-flux technique we grew superconducting $(K_{1-z}Na_z)_xFe_{2-y}Se_2$ ($z = 0.3$) single crystals. The EDX mapping revealed the uniform elements distribution on the crystal surface while the XRD measurements indicate that the crystals are compositionally inhomogenous on nanoscale. The physical properties of the as-prepared sample are characterized by electrical resistivity, magnetization and specific heat measurements. Resistivity measurements show the onset of the superconducting transition at 33 K and zero resistivity at 31.7 K. The large upper critical field $H_{c2}(0)$ was estimated as high as about of 140 T for the in-plane field and 38 T for the out-of-plane field. The anisotropy of $H_{c2}^{ab}(0)/H_{c2}^{c}(0)$ and coherence lengths $\xi^{ab}(0)/\xi^{c}(0)$ was found to be around 3.7. The pioneer studies by multiple Andreev reflections effect spectroscopy ("break-junction" technique) revealed the presence of two anisotropic superconducting gaps $\Delta_L = (9.3 \pm 1.5)$ meV, $\Delta_S = (1.9 \pm 0.4)$ meV, and provided measuring of the $\Delta_L(T)$ temperature dependence. The BCS-ratio for the large gap $2\Delta_L/k_BT_c^{bulk} \approx 6.3$ points to a strong electron-boson coupling in the "driving" condensate characterized by $\Delta_L$ order parameter.

I. INTRODUCTION

The discovery of iron pnictides with superconducting transition temperatures up to 56 K has aroused a new wave of research in this fascinating field leading to the emergence of new families of iron-based superconducting materials bearing anti-fluorite ($Fe_2As_2$) or ($Fe_2Ch_2$) ($Ch$ = S, Se, Te) layers. Recent studies of $A_xFe_{2-y}Se_2$ superconductors ($A$ = K, Rb, Cs, K/Tl, Rb/Tl) with $T_c$ of about 30 K discovered in 2010 [1] revealed that this family possesses a number of electronic and structural features significantly different from those observed in other families of iron-based superconductors.

Band-structure calculations [2] showed Fe-3$d$ bands crossing the Fermi level, thus implying metallic-type conductivity, and the density of states (DOS) at $E_F$ formed mainly by Fe-3$d$ states. This results in two quasi-two-dimensional (2D) electron-like Fermi surface sheets around M point and 3D electron-like pockets around Z point of the Brillouin zone. Unlike other Fe-based pnictides and chalcogenides, hole-like sheets are absent in the stoichiometric compound [2-4], calling into question the possibility of $s^{\pm}$-type of pairing in superconducting state [2], or even by $d$-wave symmetry of the order parameter [3, 5]. However, the hole-like pockets were shown to arise near Γ point under strong hole doping of about 60% [6]. The simple BCS-like estimates [6] based on the experimental $T_c$ values correlate with the total DOS at $E_F$, like in pnictides [6].

The absence of hole-like bands crossing the Fermi level was confirmed by angle-resolved photoemission spectroscopy (ARPES) measurements [8-11]. Moreover, the isotropic nature of the nodeless SC gaps is widely reported. A single isotropic gap opening at electron-like sheets around M point was observed in [12], with the value of BCS ratio $2\Delta^M/k_BT_c^{bulk} \approx 6.5$, and 7.4, respectively, whereas [11] points to a smaller isotropic gap ($2\Delta^Z/k_BT_c^{bulk} \approx 6$) located at the electron-band around Z point. At the same time, Zhang et al. [10] detected both superconducting gaps together (with close BCS ratio values, $2\Delta^M/k_BT_c^{bulk} \approx 8$ and $2\Delta^Z/k_BT_c^{bulk} \approx 5.4$, respectively).

Two possible gap topologies were found from the temperature dependence of a spin-lattice relaxation rate obtained in $^{77}$Se-nuclear magnetic resonance (NMR) measurements [13]: the presence of two s-wave gaps (with $2\Delta/k_BT_c^{bulk} \approx 7.2$, and 3.2, respectively) or a single s-wave gap ($2\Delta/k_BT_c^{bulk} \approx 8$). Although the results of [13] cannot distinguish between the aforementioned models, two nodeless isotropic gap values calculated are compatible with the ARPES data [10-12]. Specific heat measurements on $K_xFe_{2-y}Se_2$ single crystals ($T_c^{bulk} \approx 28$ K) also demonstrated a presence of a nodeless superconducting gap [14].

Two distinct gaps opening below $T_c^{local} = 28$ K was confirmed by scanning tunneling microscopy (STM) on $K_xFe_{2-y}Se_2$ film [15, 16]. The large gap $\Delta = 4$ meV obtained in stoichiometric $KFe_2Se_2$ phase was two times smaller than the ARPES data [10, 12], which could be explained by the gap inhomogeneity at the sample surface or by the presence of Fe vacations [15]. On the other hand, STM detected also a smaller gap of about 1 meV, which is irresolvable by ARPES. It should be also noted that in the regions of the so-called $\sqrt{2} \times \sqrt{5}$ superconducting phase a single nearly isotropic gap $\Delta = 8.8$ meV was found by STM [16]. The detailed review of theoretical and experimental studies of $A_xFe_{2-y}Se_2$ compounds was given in [4].

An important role of Fe as well as alkali metal vacancies which could produce the charge carrier doping [17] and yield changes in microstructure such as phase separation and local structure distortions [18] should be additionally stressed for $A_xFe_{2-y}Se_2$ systems. Thus, $K_xFe_{2-y}Se_2$ compound was found to be phase separated into antiferromagnetic insulating and superconducting phases [19-23]. Through detailed STM [24, 25] and TEM analysis [26, 27], non-superconducting and superconducting phases can be characterized by structural modulations with the wave vector $q_1 = 1/5(a_s + 3b_s)$ for the AFM ordered regions and $q_2 = 1/2(a_s + b_s)$ for the regions with metallic behavior associated with superconductivity. Generally, the AFM ordered phase serves as a matrix stabilizing the superconducting state. However, the chemical nature and origin of the two separated phases remain unclear. Due to the intrinsic multiphase nature of the iron selenides, which behave as phase separated nanocomposites, an availability of high quality single crystals is vital since the pulverization may lead to unpredictable and non-trackable microstructural changes. The complex microstructure makes it difficult to study the effect of doping on the properties of the superconducting phase. Very recently it was shown that Co and Mn dopants have distinct differences in occupancy and hence in the mechanism of superconductivity suppression upon doping of Fe sites [28, 29]. Taking into account a high sensibility of the superconducting phase to modifications in Fe sublattice, it is important to find a kind of substitution which allows increasing the superconducting volume fraction and to enhance the stability of the superconducting phase in the sample. For this reason, substitution in the alkali metal sublattice, for example by an alkali metal-like element such as Tl, is believed to be rather attractive. It was shown by Wang et al [30, 31] that the systematic changes in the Rb/Tl ratio in (Rb$_{1-z}$Tl$_z$)$_x$Fe$_{2-y}$Se$_2$ series have no significant effect on the unit cell parameters as well as on the T$_c$ that nevertheless can be explained by the fact that this substitution cannot produce any "chemical pressure" because the radii of 8-coordinated Rb(I) and Tl(I) are almost equal. These findings call for further study of the case when the cations have significantly different ionic radii, for example, Na and K. For pnictides, it was shown that up to 50% of potassium can be successfully substituted by sodium in K$_{1-x}$Na$_x$Fe$_2$As$_2$ solid solutions, which leads to a considerable change in the cell parameters as well as in the low-temperature behavior in the series of these compounds [32, 33].

In this study we report on the successful growth and thorough studies of superconducting (K$_{1-z}$Na$_z$)$_x$Fe$_{2-y}$Se$_2$ (z = 0.3) single crystals, which represent a first example of a sodium-substituted superconducting iron selenide obtained by a conventional high temperature technique. In addition, we present multiple Andreev reflections effect spectroscopy (MARE) studies of superconducting order parameters in (K$_{1-z}$Na$_z$)$_x$Fe$_{2-y}$Se$_2$ that give direct evidence on the presence of two independent superconducting gaps $\Delta_L \approx 9.3$ meV and $\Delta_S \approx 1.9$ meV. The large gap BCS-ratio $2\Delta_L/k_BT_C^{bulk} \approx 6.3$ is close to that for Mg$_{1-x}$Al$_x$B$_2$. The significant anisotropy of the large and the small gap was detected.

## II. EXPERIMENTAL

All preparation steps were performed in an argon-filled glove box with O$_2$ and H$_2$O content less than 0.1 ppm. At first, a starting material Fe$_{1.02}$Se was obtained by reacting Fe powder (99.99%) with Se powder (99.9%) in a molar ratio Fe:Se = 1.02: 1 in a sealed quartz tube at 420∘C for 10 days with an intermediate regrinding. Then, prereacted Fe$_{1.02}$Se powder and pieces of metallic K (99.9%) or Na (99.8%) were put into a quartz tube in a ratio $A$ : Fe$_{1.02}$Se = 0.8 : 2 ($A$ = Na, K). The quartz tube was sealed under vacuum and annealed at 380$^o$C for 6 hours. The obtained products with a nominal composition $A_{0.8}$(Fe$_{1.02}$Se)$_2$ were thoroughly ground in an agate mortar. The (K$_{1-z}$Na$_z$)$_x$Fe$_{2-y}$Se$_2$ single crystals were grown by self-flux method. To achieve maximal homogeneity of K/Na distribution we used Na$_{0.8}$(Fe$_{1.02}$Se)$_2$ and K$_{0.8}$(Fe$_{1.02}$Se)$_2$ precursors in a molar ratio 3:7, respectively. The reaction mixture was put into an alumina crucible inside a small quartz tube. The small quartz tube was sealed under high vacuum, and then was placed into a subsequently evacuated and sealed larger quartz tube. The tube was heated up to 1030 $^o$C in 5 hours, kept at this temperature for 2 hours, and cooled down slowly to 700$^o$C with a rate of 6∘C/hour, following by water-quenching.

The actual composition of the crystals was determined using X-ray energy dispersive spectrometer (INCA X-sight, Oxford Instruments) mounted on a field emission scanning electron microscope JEOL JSM 6490 LV with W-cathode. Quantitative analysis of the spectra was performed using the INCA software (Oxford Instruments).

X-ray powder diffraction data were collected using a Phillips X'Pert Pro diffractometer with $CoK_\alpha$-radiation in the reflection mode. The preliminary powdered sample was placed in a vacuum-chamber during the data collection to prevent oxidation. Profile analysis including LeBail decomposition was performed using Jana2006 software [34].

The magnetization was measured using a superconducting quantum interference device 5 T magnetometer of "Quantum Design" (MPMS) and an induction AC-magnetometer with an approximately 10 Oe AC magnetic field at frequency 120 Hz. The crystals for magnetic susceptibility measurements were sealed in the thin-walled quartz capillaries. The AC susceptibility was measured for the crystal with dimensions of approximately 6 mm in length, 3 mm in width and 0.2 mm in thickness. The long side of the crystal was roughly aligned along the AC magnetic field direction to reduce the demagnetizing factor. The initial cooling of the samples was performed after switching off the AC magnetic field and the measurements were performed on heating.

Resistivity measurements were carried out within the temperature range (10 – 300) K and in the DC magnetic field up to 9 T using a "Quantum Design" physical properties measuring system (PPMS). The measurements were done using the four probe method for the two crystals cleaved out from the same larger crystal with the current flowing in the *ab*-plane. Sample dimensions were measured with a Zeiss Stemi 2000-C stereo microscope. The first crystal was 2 mm in length (0.55 mm distance between potential contacts), 0.3 mm in width and 0.045 mm in thickness, whereas the second one was 2.2 mm in length (0.55 mm distance between potential contacts), 0.4 mm in width and 0.04 mm in thickness. The electrical contacts were attached with a silver epoxy and an In-Ga alloy. During the preparation and mounting procedures the crystals were exposed to the atmosphere for approximately 10–15 minutes.

To carry out the specific heat measurements the PPMS was used. During the heat capacity measurements, the sample was cooled to the lowest temperature with an applied magnetic field (FC) and the specific heat data were obtained between 2 K and 40 K (upon warming) using the relaxation time method.

The superconducting properties were studied by Andreev spectroscopy of superconductor - constriction - superconductor (ScS) junctions [35], realized by a "break-junction" technique [36]. In order to prevent the material from decomposition in open air, the sample mounting was done in argon atmosphere. Our single crystal (thin plates of about $2\times1\times0.1$ mm$^3$) was attached to a spring sample holder by a liquid In-Ga alloy (using 4-contact connection; *ab*-plane parallel to the sample holder) and cooled down to T = 4.2 K. Subsequent gentle mechanical curving of the holder causes a microcrack generation in the layered sample, allowing its exfoliation along the *ab*-planes, and thus creating a symmetrical contact formed by two superconducting clefts separated by a weak link (constriction). Importantly, the microcrack location deep in the bulk of the sample obstructs an impurity penetration, which retains cryogenic clefts to be as clean, as possible. Therefore, the latter ensures us to avoid observation of spurious superconducting gaps arising from the surface defects, if any. Due to single crystal nature of the samples and the specific geometry set in our experiment, the current passes through the constriction along the *c*-direction. Spring-holder precise bending gives us ability to readjust the contact point on the cryogenic cleft. Since the $(K_{1-z}Na_z)_xFe_{2-y}Se_2$ samples contain the superconducting phase in the non-superconducting matrix, the "break-junction" technique could be used for a superconducting properties study.

Current-voltage characteristic (CVC) and its derivative dI(V)/dV for tunnel junction can give us valuable information about the magnitude of the superconducting gap. Multiple Andreev reflections effect [35] occurring in ballistic contact of diameter *a* less than the quasiparticle mean free path *l* [37] which represents the SnS-interface (n = normal metal) leads to an excess current at low bias voltages in CVC and a subharmonic gap structure (SGS) in the dI(V)/dV-spectrum [38]. In the case of high transparency of the n-type constriction that is typical for our "break-junction" contacts, SGS represents a sequence of dynamic conductance dips at specific bias voltages $V_n = 2\Delta/en$, where $\Delta$ is the required superconducting gap value, *e* – elementary charge, and n = 1, 2,… – subharmonic order. Therefore, using the positions of the gap peculiarities observed, we can determine the gap value within the whole range of temperatures ($0 < T \leq T_c$) directly from the experimental spectrum, *i.e.* without dynamic conductance fitting [38]. In the case of two-gap superconductor, two such SGSs should be observed. In the present study, the CVCs and dynamic conductance spectra for SnS-junctions were measured by a digital set-up controlled by the National Instruments digital board [39].

## III. RESULTS AND DISCUSSION

### A. Composition and morphology of $(K_{1-z}Na_z)_xFe_{2-y}Se_2$ single crystals

The synthesis resulted in the plate-like crystals with a shiny bronze surface grown on top of a batch and a flux consisted mainly of unreacted t-FeSe and reduced α–Fe. $(K_{1-z}Na_z)_xFe_{2-y}Se_2$ crystals grown in a layered morphology are easily cleaved along the *ab*-plane. The typical SEM image of a freshly cleaved surface perpendicular to the *c*-axis is shown in the inset to Fig. 1. The EDX analysis yields the formula of the crystal with the nominal composition $(K_{0.7}Na_{0.3})_{0.8}Fe_{2.04}Se_2$ to be $(K_{0.68(2)}Na_{0.32(2)})_{0.95(4)}Fe_{1.75(2)}Se_2$. The composition is determined by averaging over 18 points of the same specimen on the freshly cleaved crystal surface and for several crystals from the batch. One should take into account that different crystals from the same batch denoted further KNFS1, KNFS2, etc, may be compositionally inhomogenous on nanoscale, but the average composition determined by EDX is the same for all crystals. The main panel of Fig. 1 shows a typical EDX spectrum. The Na $K_\alpha$-line at 1.04 keV is clearly seen. The distribution curves of all elements given in Fig. 2 (a) do not reveal any features associated with microscopic inhomogeneities in the element distribution. EDX mapping shown in Fig. 2 (b-e) provides in addition to the conventional SEM image a meaningful picture of the element distribution of the surface. The mapping was done on the sample surface with dimensions about 100×200 μm$^2$. Clearly, EDX analysis revealed the presence of uniformly distributed Na, K, Fe and Se, suggesting that the surface of the sample is compositionally homogenous, at least within the spatial resolution of SEM-EDX analysis. Our observations allow us to state that Na does not form any separate phase but rather occupy the sites in the lattice of the $K_xFe_{2-y}Se_2$ parent phase. It should be stressed, however, that our EDX data are insufficient to determine whether Na is uniformly distributed between the superconducting and AFM phases on nanoscale or concentrates mainly in one of them.

X-ray powder diffraction pattern for the sample under consideration is given in Fig. 3. However, the interpretation of the obtained XRD data is complicated by the fact that the intrinsic multiphase nature of the compounds and their relatively low crystallinity lead to a significant broadening of the Bragg peaks. In the $(K_{1-z}Na_z)_xFe_{2-y}Se_2$ compound the presence of at least two phases was shown. The main phase can be indexed in the well-known tetragonal body-centered supercell of the original $ThCr_2Si_2$ subcell with the space group *I4/m*. The second phase described by the space group *I4/mmm* could be a variant of vacancy disordered structure. The refinement converged with $R_{B(obs)} = 2.90\ \%$, $wR_{B(obs)} = 3.71\ \%$ and GOF = 0.95.

The refined unit cells parameters of the $(K_{1-z}Na_z)_xFe_{2-y}Se_2$ sample obtained by a full-profile LeBail decomposition are $a = 8.698(1)$ Å, $c = 14.108(2)$ Å for the main phase and $a = 3.946(2)$ Å, $c = 14.302(7)$ Å for the minor phase. Thus, sodium doping significantly decreases $a$ parameter of the main phase in comparison with $K_xFe_{2-y}Se_2$ compound [40, 41, 42] whereas $c$ parameter remains essentially the same. For the minor phase associated with superconductivity the opposite tendency is observed. Since the incorporation of Na induces the structural changes in the major as well as in the minor phase, one may conclude that sodium is presented in both phases, though, perhaps, in different amounts.

### B. Magnetic susceptibility measurements

Fig. 4 shows the temperature dependences of the real and the imaginary parts of the susceptibility for the typical $(K_{1-z}Na_z)_xFe_{2-y}Se_2$ sample. The absolute value of susceptibility saturates to -0.96 at 20 K reflecting a nearly complete diamagnetic screening. A small difference between the measured Meissner screening volume and the sample volume may originate from a partial misalignment of the sample.

### C. Resistivity measurements

Fig. 5 shows temperature dependences of resistivity for KNFS1 and KNFS2 crystals together with a magnified region around the superconducting transition and low temperature parts of the dependences plotted versus $T^3$ on two insets. The ratios of the maximum value of resistivity and the minimum value of normal resistivity are in the range 15–20 reflecting the high quality of the crystals. The absolute values of resistivity and the overall shape of R(T) curves are similar to those reported for $K_xFe_{2-y}Se_2$ [43, 44] and $Rb_xFe_{2-y}Se_2$ [45]. However, it is important to stress that the absolute values of resistivity are two or three orders higher than the corresponding values for LiFeAs [46] and doped Ba122 [47] systems. Another apparent difference from the mentioned pnictide superconductors is a higher value of exponent in the power law approximation for resistivity at low temperatures. Quadratic temperature dependence of resistivity for 111 and 122 pnictides is considered as a manifestation of the strong electron-electron correlation [46, 47] in these compounds. For our samples the temperature dependence of resistivity is rather cubic in 60–100 K range with a crossover to a higher power at lower temperatures (see the upper inset in Fig. 5). Such a behavior may imply a predominance of the spin orbital scattering in these compounds. On the other hand, this behavior can reflect a saturation of resistivity at the relatively high temperature due to a peculiarity of the samples microstructure. The difference in resistivity of the two studied crystals exceeds the possible error due to the limited accuracy of the geometric calculations. Moreover, maximum and minimum values of the normal resistivity are not scaled. Since these two samples are the parts of the same crystal, it can mean a noticeable inhomogeneity of the crystals on nanoscales.

The resistivity of the $(K_{1-z}Na_z)_xFe_{2-y}Se_2$ crystals can be fitted using a model of two percolating phases that act as resistors in parallel, one with a metallic Bloch-Grüneisen temperature dependence $\rho_{metal}(T) = \rho_{m0} + AT^n$, and the other one with a Boltzmann-type insulating temperature dependence $\rho_{semim}(T) = \rho_{sm0} \exp(E_g / 2k_BT)$. The fit given in Fig. 5 allows us to estimate $n$ to be $2.95 \pm 0.1$, and the insulator activation energy $E_g = (76 \pm 2)$ meV which scales well with the results obtained in [40].

Despite the difference in the normal resistivity, other studied properties of the crystals including superconducting properties are very similar. The samples show the same positive transverse magnetoresistance which is proportional to the square of the field (Fig. 6). Both crystals show very sharp and perfectly coinciding superconducting transitions at 32.5 K (onset) as it seen in the lower inset in Fig. 5. The dependences of the transition temperatures on the magnetic field are also very similar.

The data of $H_{c2}(T)$ at 90% and 50% resistive transition for $H//c$ and $H//ab$ are plotted in Fig. 7. The curves for 90% deviate appreciably from the linearity. Linear fits in the field range from 0 to 3 T give $-dH_{c2}^c/dT = 3.4$ T/K and $-dH_{c2}^{ab}/dT = 24.8$ T/K whereas in the field range from 3 T to 9T liner fits give $-dH_{c2}^c/dT = 2.3$ T/K and $-dH_{c2}^{ab}/dT = 10.7$ T/K. Linear fits of the curves corresponding to 50% resistivity transition threshold give $-dH_{c2}^c/dT = 1.7$ T/K and $-dH_{c2}^{ab}/dT = 6.3$ T/K. The rough estimation of $H_{c2} = -0.69\ T_c\ dH_{c2}/dT$ [48] gives $H_{c2}^c \approx 38$ T and $H_{c2}^{ab} \approx 140$ T and allows assessing the coherence length $\xi$: $\xi^{ab} = (330\text{ nm}^2/\text{T} / H_{c2}^c)^{1/2} \approx 2.9$ nm; $\xi^c = 330\text{ nm}^2/\text{T} / (H_{c2}^{ab}\ \xi^{ab}) \approx 0.8$ nm. The last values are close to the corresponding values reported for $K_xFe_{2-y}Se_2$ [43] and for $Rb_xFe_{2-y}Se_2$ [49]. The anisotropy of $H_{c2}^{ab}(0)/H_{c2}^c(0)$ and $\xi^{ab}(0)/\xi^c(0)$ is found to be around 3.7.

### D. Specific heat measurements

A pronounced jump due to the superconducting transition can be observed in the temperature dependence of zero field specific heat for the $(K_{1-z}Na_z)_xFe_{2-y}Se_2$ sample, as shown in Fig. 8. The normal-state specific heat can be described by the equation $C_p(T,H) = C_{el} + C_{lattice} = \gamma_n T + \beta T^3 + \eta T^5$, where $\gamma_n T$ is the electron contribution in the heat capacity, $\beta T^3 + \eta T^5$ is the phonon part of heat capacity. The solid red line is the best fit of $C_p/T$ data above $T_c$ yielding $\gamma_n \approx 5.3$ mJ/mol·K$^2$, $\beta \approx 1.27$ mJ/mol·K$^4$. Using the obtained value of $\beta$ and the relation $\theta_D = (12\pi^4 k_B N_A Z/5\beta)^{1/3}$, where $N_A = 6.02 \times 10^{23}$ mol$^{-1}$ is the Avogadro constant, $k_B = 1.38 \times 10^{-23}$ J·K$^{-1}$ is the Boltzmann constant, $Z = 5$ the number of atoms per formula unit, we get the Debye temperature $\theta_D \sim 197$ K. The upper inset to Fig. 8 shows the specific heat data after subtracting both electron and phonon contributions of the normal state to the total specific heat and plotted as $(C-C_n)/T$ vs. $T$. The observed specific heat anomaly for $(K_{1-z}Na_z)_xFe_{2-y}Se_2$ is similar to that of $K_xFe_{2-y}Se_2$ and $Rb_xFe_{2-y}Se_2$ [14, 50] but smaller than the value for other FeAs-based superconductors [51-54].

The temperature dependence of the low temperature part of the specific heat data (5-13 K) was plotted as $C_p/T$ vs. $T^2$. The fact that the low temperature specific heat data show a linear behavior within a wide range of temperatures indicates the absence of Schottky anomaly (see the lower inset to Fig. 8). The value of the residual Sommerfeld coefficient $\gamma_r = 0.19$ mJ/mol·K$^2$ was determined from the fit of the experimental data. Similar values of $\gamma_r$ were reported for the related SC $Rb_xFe_{2-y}Se_2$ compounds [50]. Assuming that the residual Sommerfeld coefficient $\gamma_r$ corresponds to the fraction of the normal conducting state, the obtained ratio of $\gamma_r/\gamma_n$ implies that the volume fraction of the superconducting phase in the sample is 96%.

E. Multiple Andreev reflections effect spectroscopy

In our "break-junction" studies, SnS-Andreev contacts were realized on cryogenic clefts in KNFS1 and KNFS3 samples. The CVCs for contacts #d4, KNFS1 sample, or, briefly, KNFS1_d4, and KNFS3_d5 measured at $T = 4.2$ K are presented in Fig. 9 for comparison and marked as $I_4(V)$ and $I_5(V)$, respectively. One could detect a pronounced excess current at low bias voltages typical for Andreev transport, which allows distinguishing confidently whether the contact is in Andreev or Josephson regime. Therefore, the data presented indicate the constriction formed between two cryogenic clefts of the break-junction to act as a normal metal. The dynamic conductance spectra for the aforementioned contacts in Fig. 9 are labeled as $dI_4(V)/dV$, $dI_5(V)/dV$. The characteristics were shifted along the vertical scale for clarity; background was suppressed. The subharmonic gap structure (SGS) containing two well-defined conductance dips located at $V_1 \approx \pm(16.2 \div 20.8)$ mV and $V_2 \approx \pm(8.1 \div 10.2)$ mV (marked by $n_L = 1$, $n_L = 2$, respectively; the $n_L = 1$ labels point to the doublet centers) are clearly visible in both spectra. Using the corresponding dip positions $V_n$ for subharmonic orders $n = 1, 2$, we obtain, in accordance with formula $V_n = 2\Delta/en$ [38], the average value of the superconducting gap $\Delta^{aver} \approx 9.3$ meV for KNFS1_d4 contact, and $\Delta^{aver} \approx 8.6$ meV for KNFS3_d5. The position of the peculiarity at $V \approx \pm 3.8$ mV ($n_S = 1$ labels) does not satisfy that for the third Andreev subharmonic expected at $V_3 \approx \pm(5.4 \div 6.9)$ mV, and, therefore, may be interpreted as the first minimum for the small gap $\Delta_S$ of about 1.9 meV. Noteworthy, both peculiarities for the large gap are doublet-shaped, which could be caused by gap anisotropy in k-space. Since SnS-Andreev spectroscopy is able to reveal the gap magnitude rather than its location in k-space, the doublet structure can be described either as a pair of two distinct isotropic gaps $\Delta_L^{a,b}$ with close values opening at different Fermi surface sheets $\Delta_L^a = (10.5 \pm 1.5)$ meV, $\Delta_L^b = (8.1 \pm 1.2)$ meV (agrees well with some ARPES data [10, 11]), or as a single large gap $\Delta^{aver} \approx 9.3$ meV with up to $(23 \pm 6)\%$ anisotropy in k-space (extended s-wave symmetry).

To resolve clear Andreev dips for the small gap, we present the excess-current CVC and dynamic conductance spectrum (with exponential background suppressed) for KNFS1_d8 contact measured at T = 2.4 K (Fig. 10). Sharp doublet-minimum located at $V \approx \pm 4$ mV (labeled as $n_S = 1$) being more intensive than the second peculiarity for the large gap ($n_L = 2$) demonstrates, in accordance with the theory [38], the onset of the SGS for the small gap $\Delta_S \approx 2$ meV. The $\Delta_S$ anisotropy is about 18% (extended s-wave symmetry). The doublet is well-resolved only at low temperatures, when the temperature smearing factor is not so significant. The $n_L = 2$ minima is wide enough and has a complex fine structure due to the anisotropy. Its average position describes the large gap $\Delta_L \approx 9$ meV. Importantly, the fine structure of the large gap peculiarities (triplet) is reproduced in the SnS-spectra obtained on different $(K_{1-z}Na_z)_xFe_{2-y}Se_2$ samples.

The SGS formula implies a linear relation between the position of Andreev peculiarities $V_n$ and their inversed number, $1/n$. Such a dependence plotted for the spectra studied is shown in the inset of Fig. 10. Solid symbols depict SGS positions for the large gap $V_n$, open symbols belong to the small gap: both minima in the doublets for the KNFS1_d8 contact spectrum are marked separately. Due to the dramatic rise of the excess current when approaching $V \to 0$, and therefore increasing of the dynamic conductance, the higher order Andreev minima for the small gap become unresolved. The experimental data are fitted by two straight lines with different slopes, both essentially crossing at the (0; 0)-point. Obviously, this means reproducing SGS corresponding to the averaged value of the large gap $\Delta_L^{aver} = (9.3 \pm 1.5)$ meV. The small gap $\Delta_S^{aver} = (1.9 \pm 0.4)$ meV.

Calculating the averaged BCS-ratio for the components of the large gap doublet (see the inset of Fig. 10), we get $2\Delta_L^a/k_B T_c^{bulk} \approx 7.2$, $2\Delta_L^b/k_B T_c^{bulk} \approx 5.9$, both much more than 3.52, and conclude on a strong electron-boson coupling in the bands with $\Delta_L$ order parameter. The values obtained are in good correspondence with the ARPES data [10-12]. On the contrary, the BCS-ratio for the small gap $2\Delta_S/k_B T_c^{bulk} \approx 1.3$ does not exceed the BCS-limit.

Figure 11 shows the temperature behavior for the large gap doublet $\Delta_L^{a,b}$ (up and down triangles; plotted on the base of the KNFS1_d4 spectrum measurements within 4.2 K $\leq$ T $\leq$ 34 K), the dependence for averaged gap $\Delta_L^{aver}$ is shown by open circles. Gray rhombs represent how the $\Delta_L^a - \Delta_L^b$ difference depends on the temperature variation. Note, its behavior resembles neither that of $\Delta_L^a(T)$, nor that of $\Delta_L^b(T)$, thus demonstrating that both Andreev peculiarities do not form an SGS, but have the same subharmonic order *n*. The abrupt closing of the $\Delta_L$ at $T_c^{local} \approx 32.5$ K which corresponds to the contact area (usually less than 100 nm in diameter) transition to the normal state, agrees on the whole with the standard single-band BCS-like function (dashed lines). However, the $\Delta_L^{aver}(T)$ dependence slightly bends down from the BCS-like curve. Importantly, such a deviation observed is typical for the "driving" gap temperature dependence in the two-gap BCS-model suggested in [55, 56] due to the interband interactions between $\Delta_L$-condensate(s) and the condensate characterized by the small gap $\Delta_S$, and was experimentally observed earlier on two-gap superconductors [57-61]. On this basis, we can also suppose an at least two-gap superconductivity scenario in $(K_{1-z}Na_z)_xFe_{2-y}Se_2$.

The large gap BCS-ratios $2\Delta_L^{aver}/k_B T_c^{local} \approx 6.6$ and $2\Delta_S/k_B T_c^{local} \approx 1.3$ for the small gap agree with the values averaged over the number of iron-pnictide samples presented in [62]. The reduced BCS-ratio value for the small gap can be a consequence of an induced superconductivity in $\Delta_S$-bands at temperatures from T $\approx$ 7.5 K (where the $\Delta_L(T)$ starts to deviate from the single-gap function, see Fig. 11) up to $T_c^{local}$ due to a k-space proximity effect between the "driving" and the "driven" condensates that essentially means nonzero interband coupling constants. In addition, the bulk critical temperature for the sample KNFS1 (see R(T) superconducting transition in Fig. 11), being one of the highest among the samples synthesized, nearly coincide with the $T_c^{local}$ for the junction.

IV. CONCLUSIONS

In summary, the crystals of mixed alkali metal $(K_{1-z}Na_z)_xFe_{2-y}Se_2$ (z = 0.3) iron selenide with the superconducting transition temperature $T_c \approx 32$ K were successfully grown using the self-flux technique. The physical properties of the as-prepared samples are characterized by electrical resistivity, magnetization and specific heat measurements. The large upper critical field $H_{c2}(0)$ was determined in the *ab*-plane and along the *c*-axis. The anisotropy of superconductivity determined by the ratio of $H_{c2}^{ab}$ and $H_{c2}^{c}$ estimated to be 3.7 is larger than that in pnictide, but smaller than that in cuprate superconductors.

Superconducting properties of $(K_{1-z}Na_z)_xFe_{2-y}Se_2$ were studied for the first time by multiple Andreev reflections effect spectroscopy (MARE). It could not provide the direct information on gap distribution in k-space, but the significant anisotropy of the large gap leads to the two following scenarios: the existence of two distinct isotropic gaps ($\Delta_L^a = (10.5 \pm 1.5)$ meV, $\Delta_L^b = (8.1 \pm 1.2)$ meV) at different Fermi surface sheets, or one extended s-wave gap $\Delta_L^{aver} \approx 9.3$ meV of about 23% anisotropy in k-space. Due to the asymmetry of Andreev peculiarities is slight, one may assert the absence of nodes in the k-space distribution of superconducting gaps. The small gap $\Delta_S = (1.9 \pm 0.4)$ meV with footprints of anisotropy was also observed. Typical bending down of the $\Delta_L(T)$ temperature dependences with respect to the single-band BCS-like behavior unambiguously points to a nonzero interband interaction between the two condensates (k-space proximity effect). The BCS-ratios calculated $2\Delta_S/k_BT_c^{local} \approx 1.3$, $2\Delta_L^{aver}/k_BT_c^{local} \approx 6.6$ (for the components of the doublet $2\Delta_L^a/k_BT_c^{bulk} \approx 7.2$, $2\Delta_L^b/k_BT_c^{bulk} \approx 5.9$), suggest a strong electron-boson coupling in $\Delta_L$-bands and proximity-induced superconductivity in $\Delta_S$-bands.

The properties of the $(K_{1-z}Na_z)_xFe_{2-y}Se_2$ samples seem to be rather similar to those of undoped potassium ferroselenide $K_xFe_{2-y}Se_2$. This may point to minor variations in superconducting phase composition under K by Na substitution, which is possible, for example, in the case of irregular sodium distribution in the coexisting phases on nanoscale revealed by our XRD measurement.

ACKNOWLEDGMENT

The authors are grateful to Prof. Ya. G. Ponomarev for providing techniques. This work was supported in part by M. V. Lomonosov Moscow State University Program of Development. Financial support by 16.120.11.3264-MK of Russian MES and RFBR 12-03-91674-ERA_a, 12-03-01143 and 12-03-31717 is cordially acknowledged.

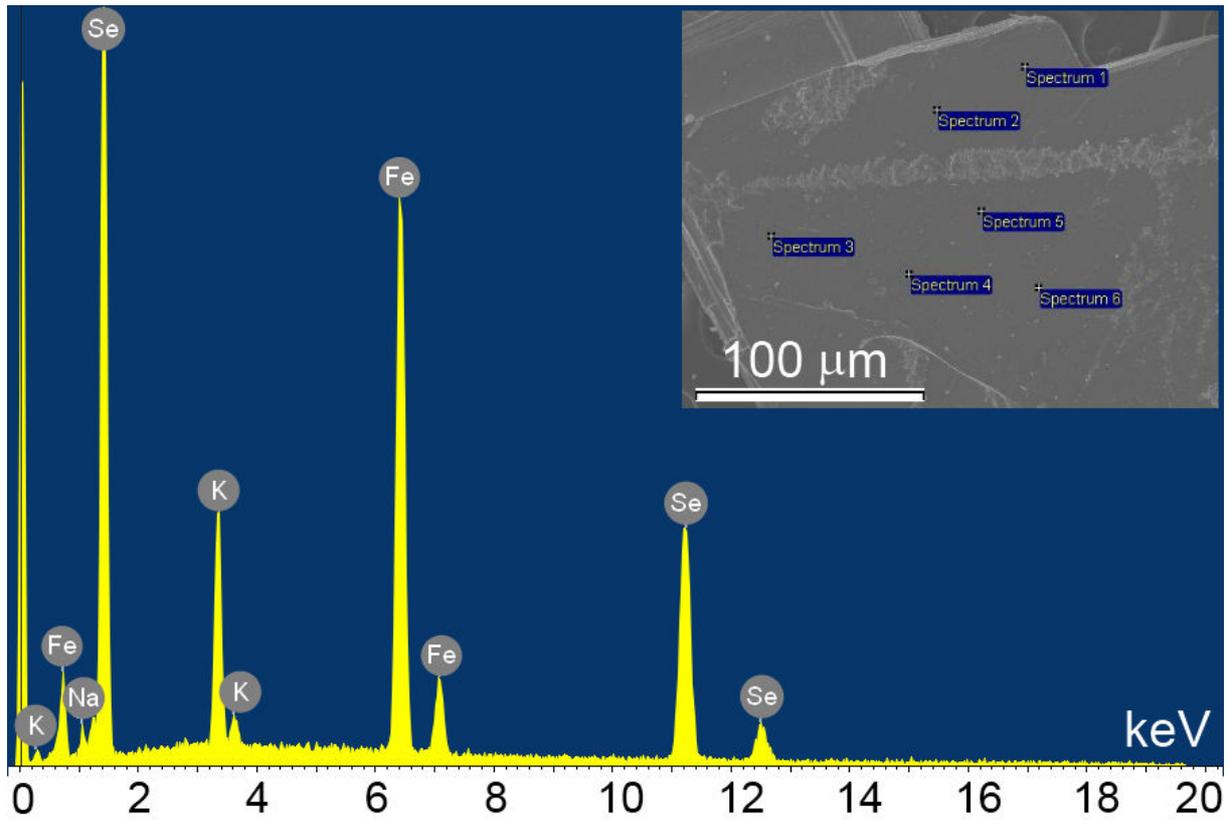

**Fig. 1**. EDX spectrum of a $(K_{1-z}Na_z)_xFe_{2-y}Se_2$ single crystal. The inset shows a SEM image of the specimen acquired at 20 keV in the secondary electron mode.

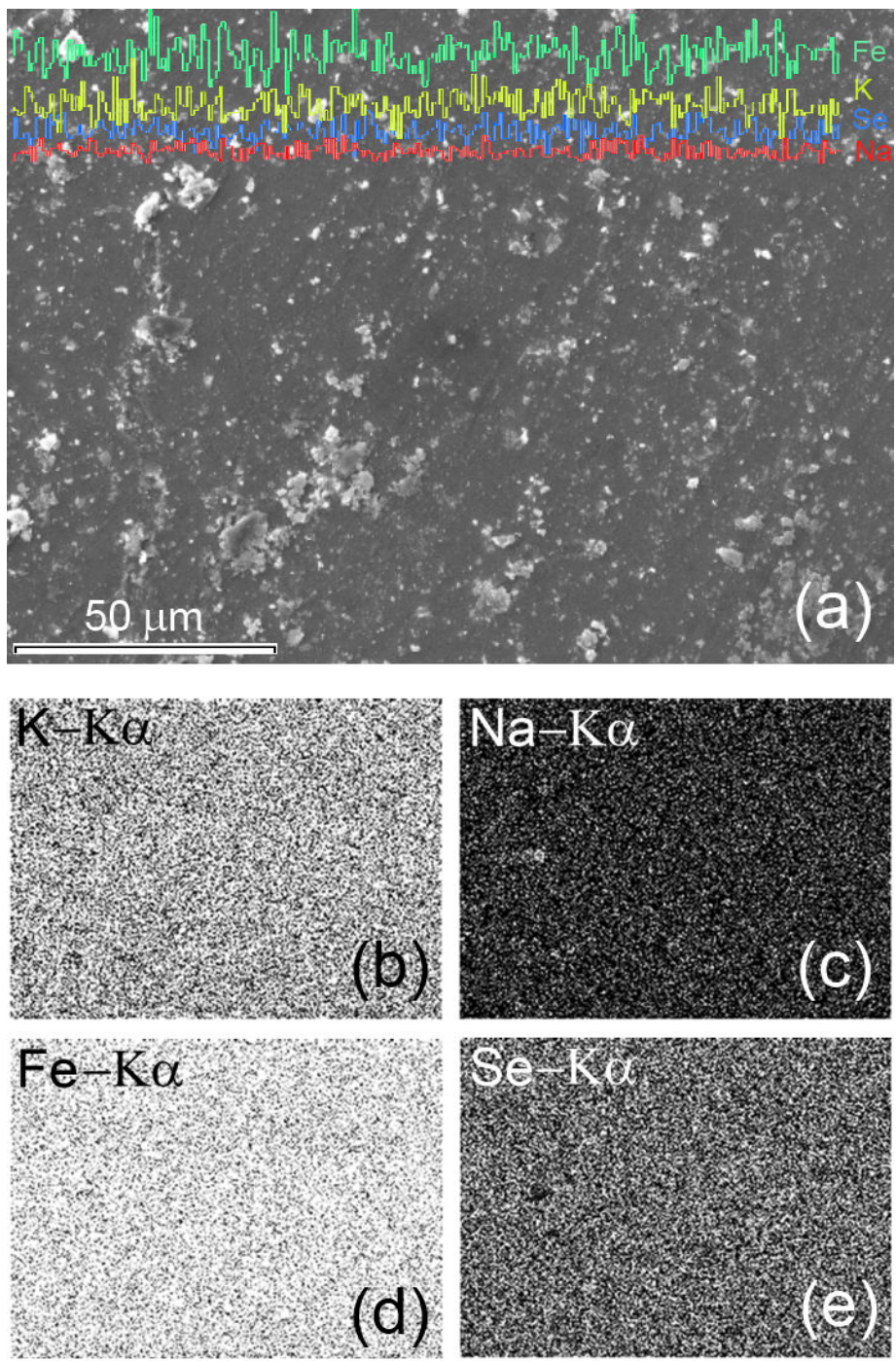

**Fig. 2**. EDX analysis of $(K_{1-z}Na_z)_xFe_{2-y}Se_2$. Panel (a) shows the SEM image of the analyzed surface. The element distribution curves along the selected direction are given in the upper part of the figure. Panels (b)-(e) show the mapping of the K, Na, Fe and Se intensity distribution.

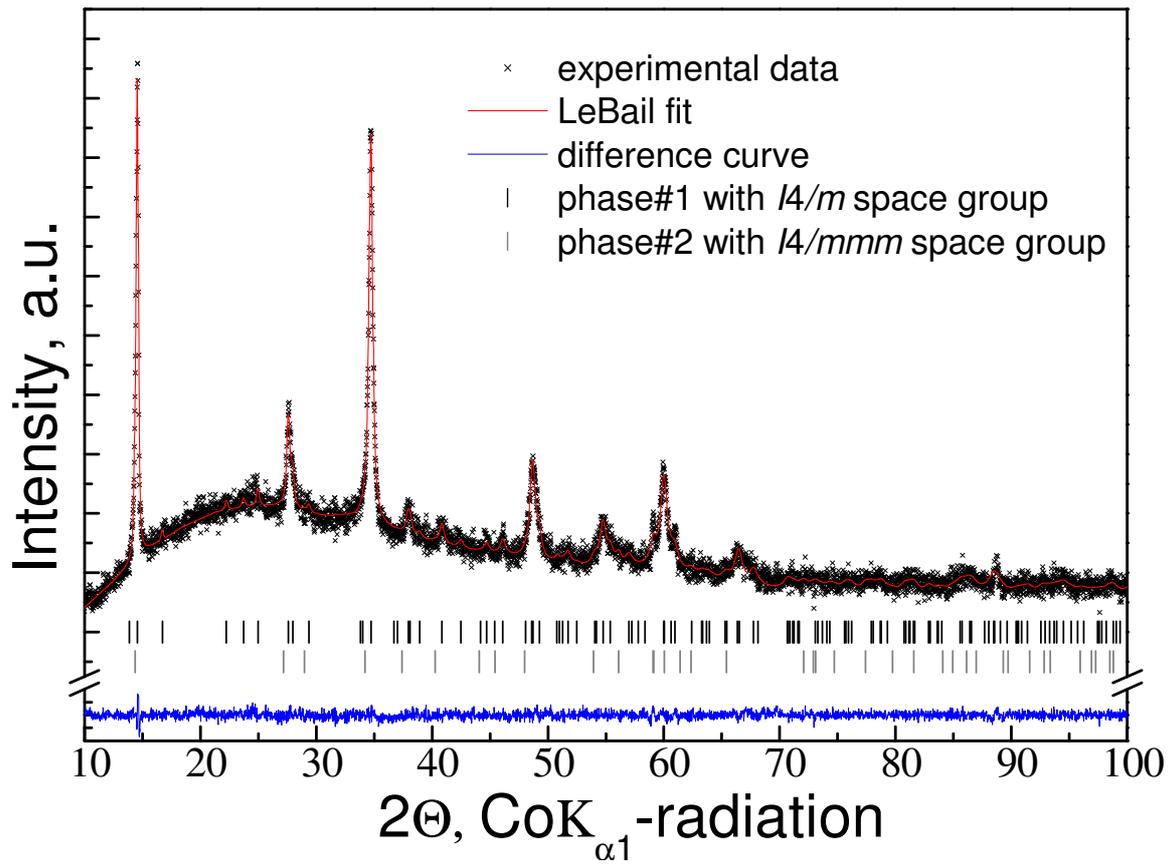

**Fig. 3**. X-ray diffraction pattern of a ground $(K_{1-z}Na_z)_xFe_{2-y}Se_2$ sample. Crosses are the experimental data, solid line is the LeBail fit, tick marks denote the positions of Bragg reflections, given below is the difference curve.

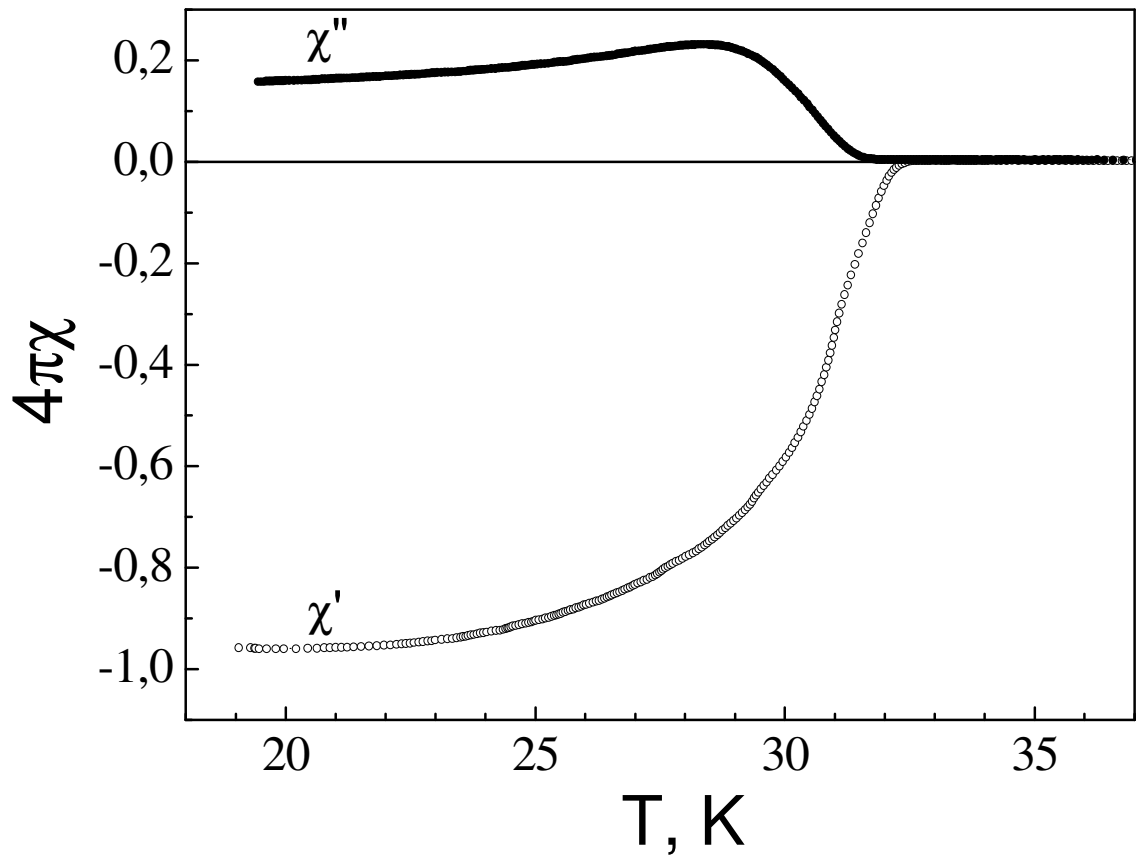

**Fig. 4**. The real and the imaginary parts of the magnetic susceptibility for the $(K_{1-z}Na_z)_xFe_{2-y}Se_2$ crystal taken at H = 10 Oe applied parallel to *ab*-plane.

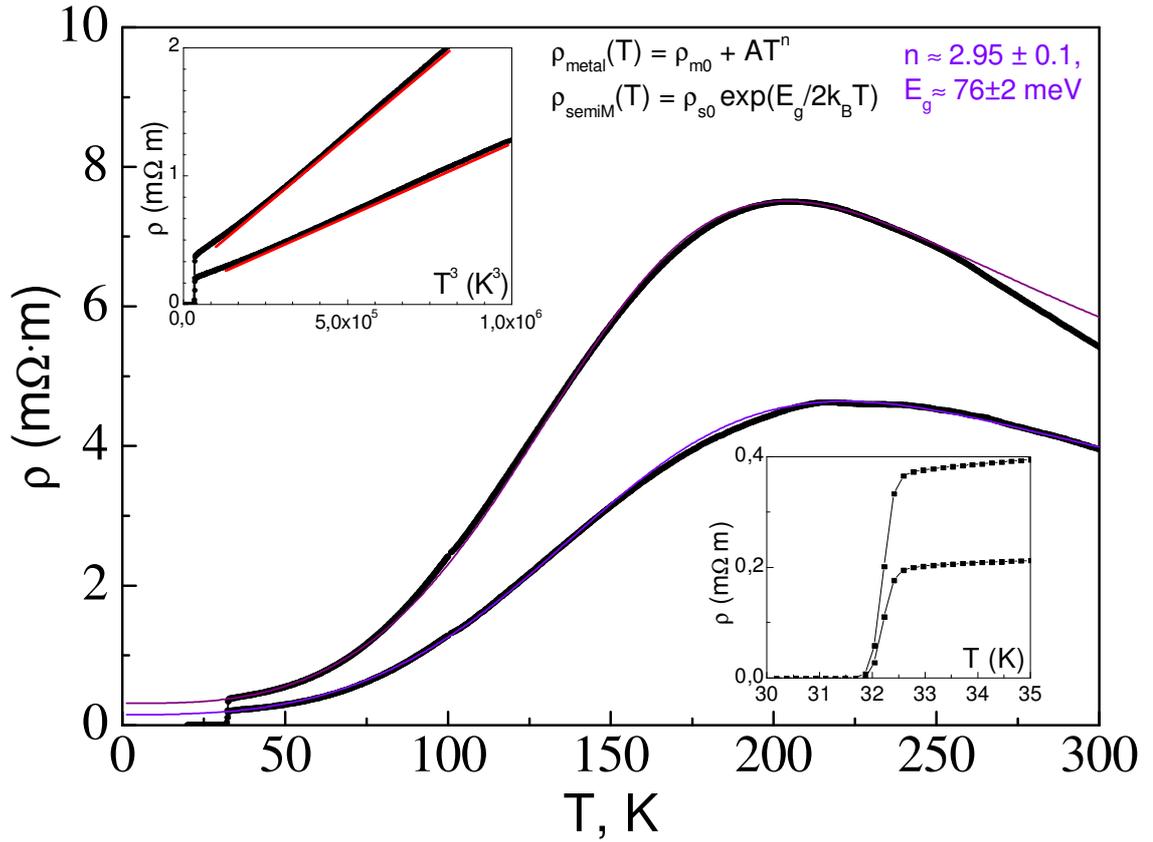

**Fig. 5**. Resistivity dependence for two $(K_{1-z}Na_z)_xFe_{2-y}Se_2$ superconducting crystals. The resistivity behavior can be fitted as a metal-insulator composite over the full temperature range (violet and purple solid lines). For both curves fitting parameters are about $n = (2.95 \pm 0.1)$, and $E_g = (76 \pm 2)$ meV. The upper inset shows the cubic temperature dependence of resistivity in (60 – 100) K range. The lower inset presents the coinciding superconducting transitions at 32.5 K for both crystals.

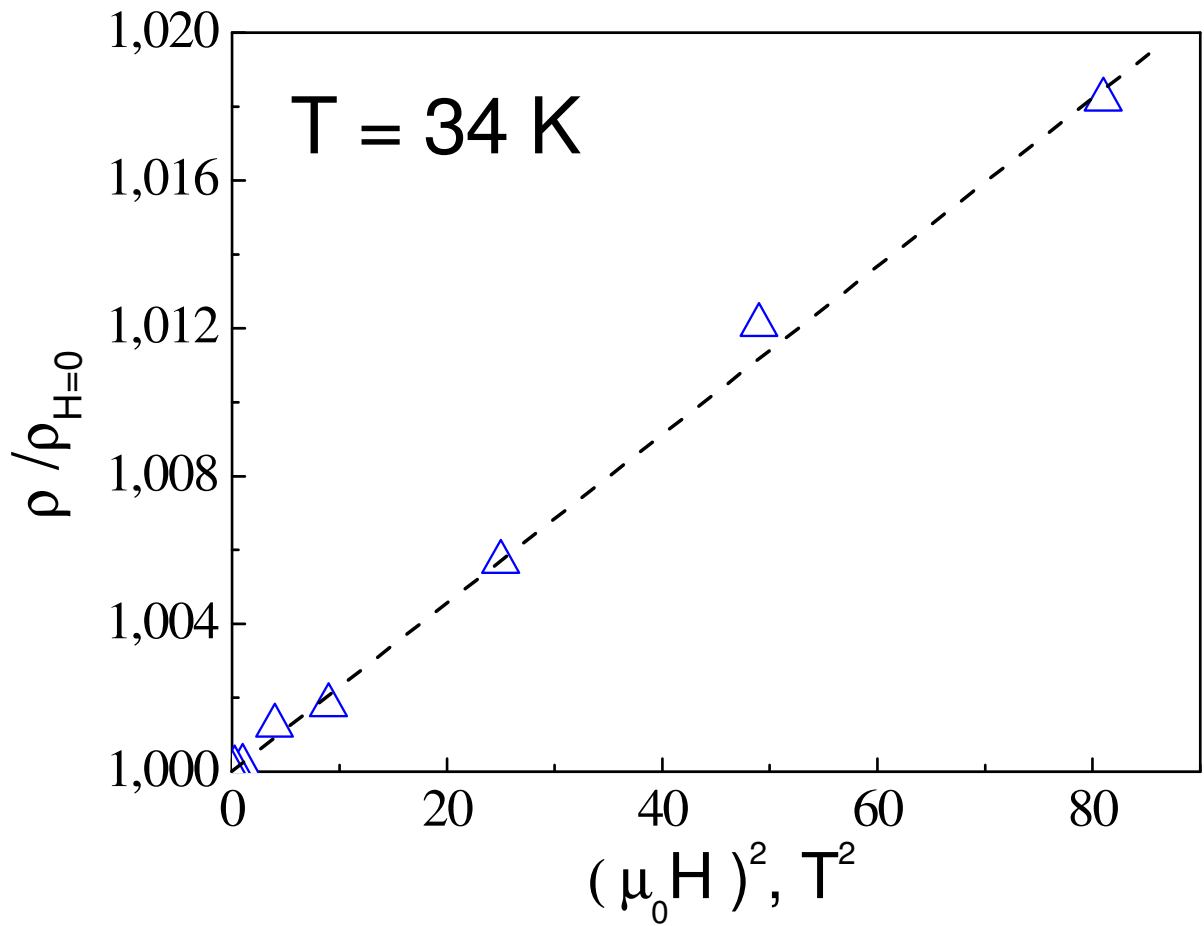

**Fig. 6**. The transverse magnetoresistance plotted as ρ/ρ$_{H=0}$ *vs*. (μ$_0$H)$^2$.

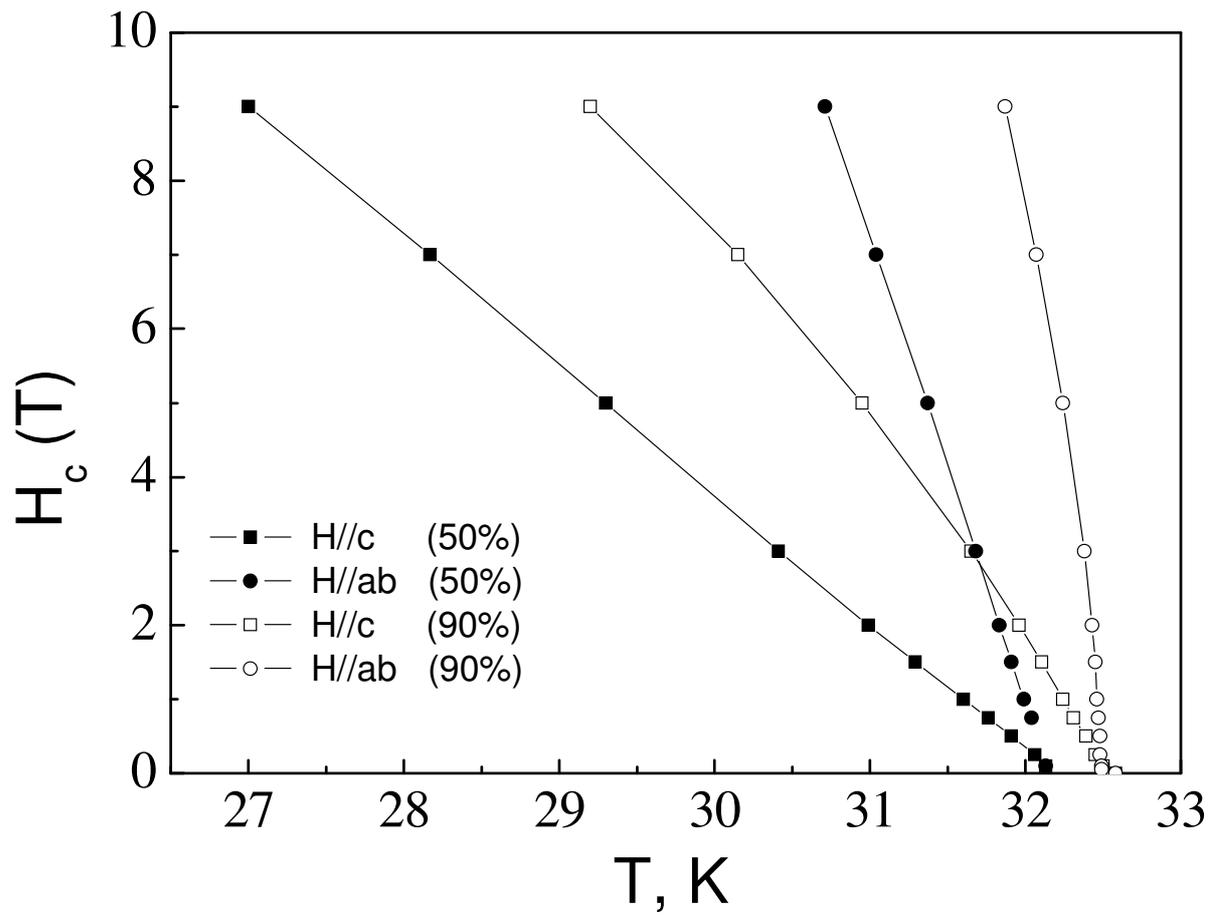

**Fig. 7**. The determination of $H_{c2}(T)$ at 90% and 50% resistive transition for the in-plane and out-of-plane fields.

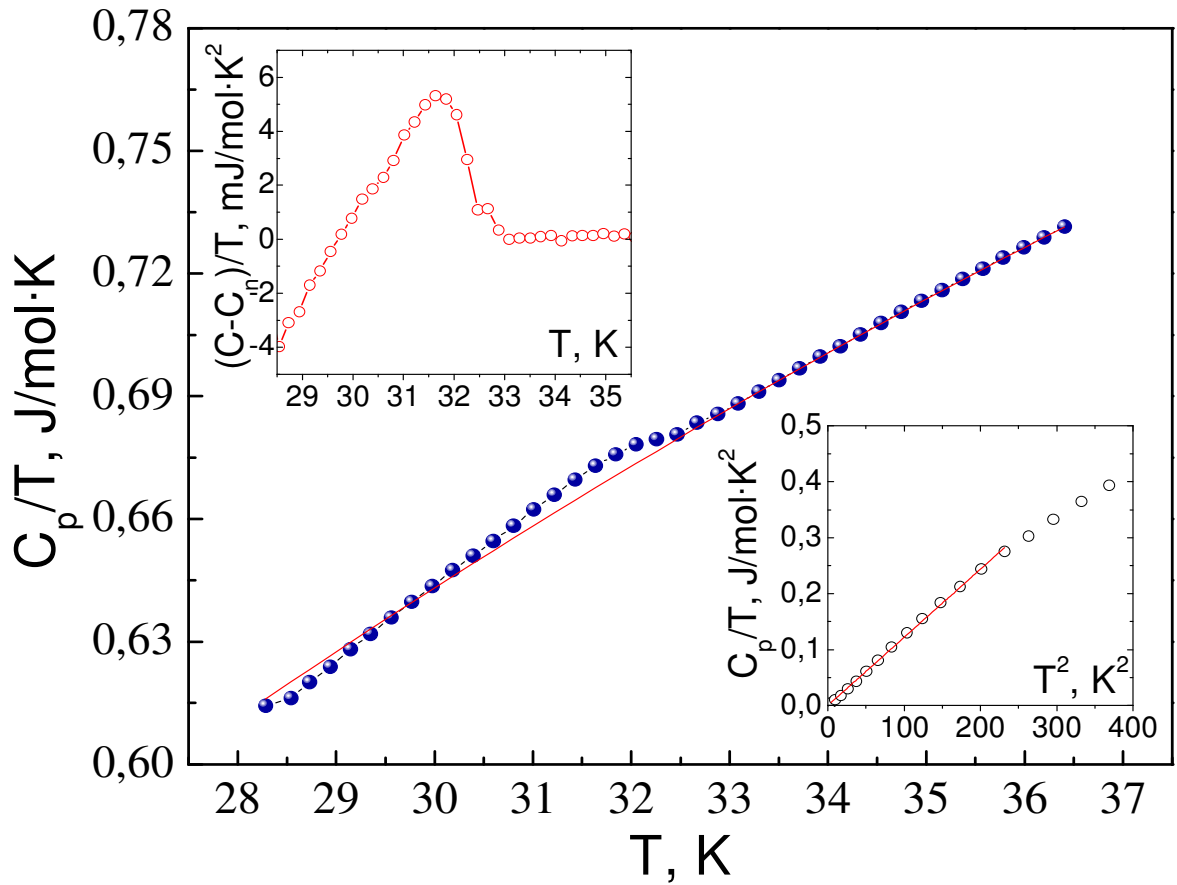

**Fig. 8**. The specific heat data near the transition temperature plotted as $C_p/T$ *vs.* T. The upper inset shows the specific heat data after subtracting both electronic and phononic contributions of the normal state to the total specific heat and plotted as $(C-C_n)/T$ *vs.* T. The lower inset shows the low-temperature region together with a fit of $C_p/T$ *vs.* $T^2$.

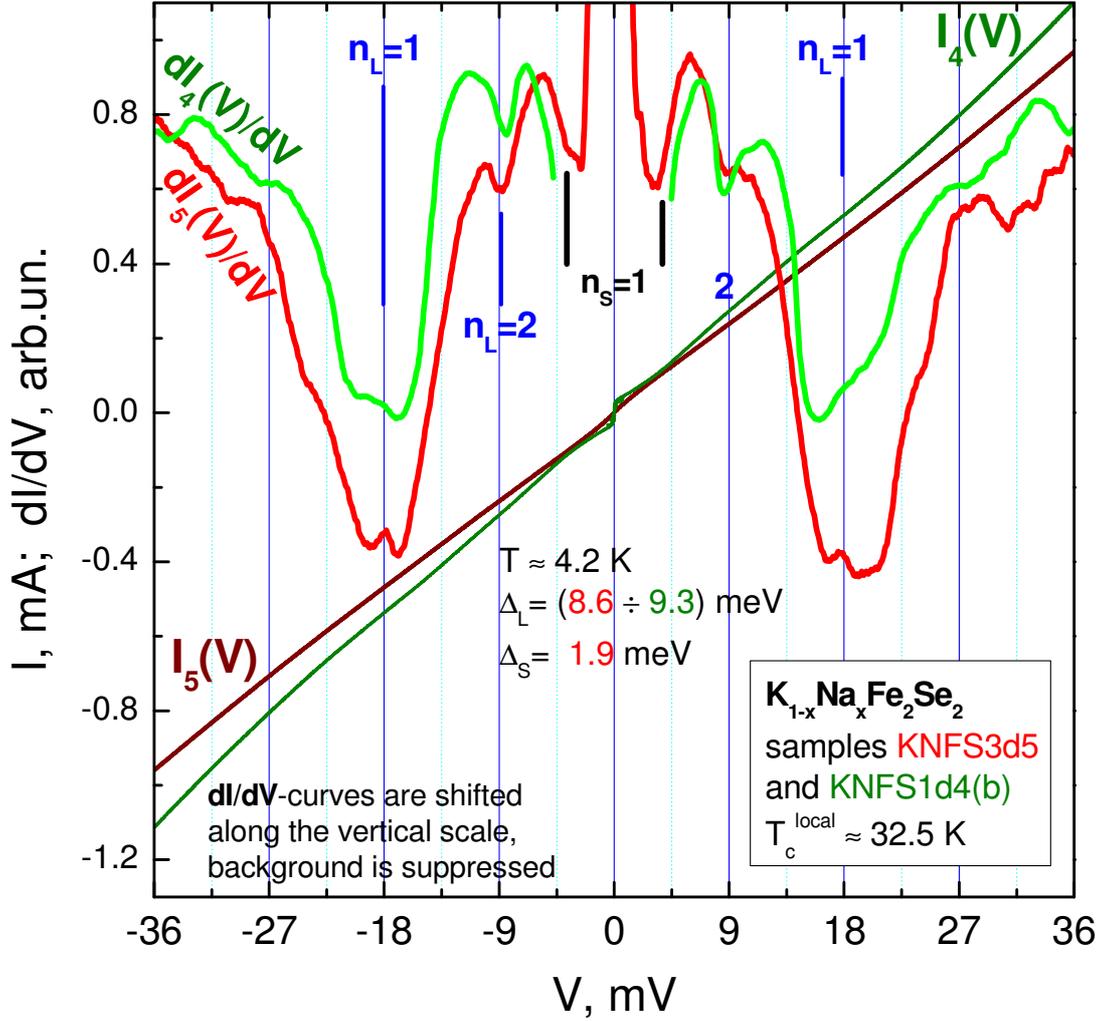

**Fig. 9.** Current-voltage characteristics (CVC; marked as $I_4(V)$, $I_5(V)$) and the dynamic conductance ($dI_4(V)/dV$, $dI_5(V)/dV$) for SnS-contacts #d4, sample KNFS1 (or, briefly, KNFS1_d4), and KNFS3_d5, respectively. The data were measured at T = 4.2 K. Local critical temperature for KNFS1_d4 contact is about 32.5 K. dI(V)/dV-spectra were shifted along the vertical scale for clarity; background was suppressed. Subharmonic gap structure (SGS) dips for the large gap $\Delta_L \approx 9.3$ meV for KNFS1_d4 contact, and $\Delta_L \approx 8.6$ meV for KNFS3_d5 are signed by $n_L=1$, $n_L=2$ labels; minimum positions for the small gap $\Delta_S = 1.9$ meV are signed by $n_S=1$ labels.

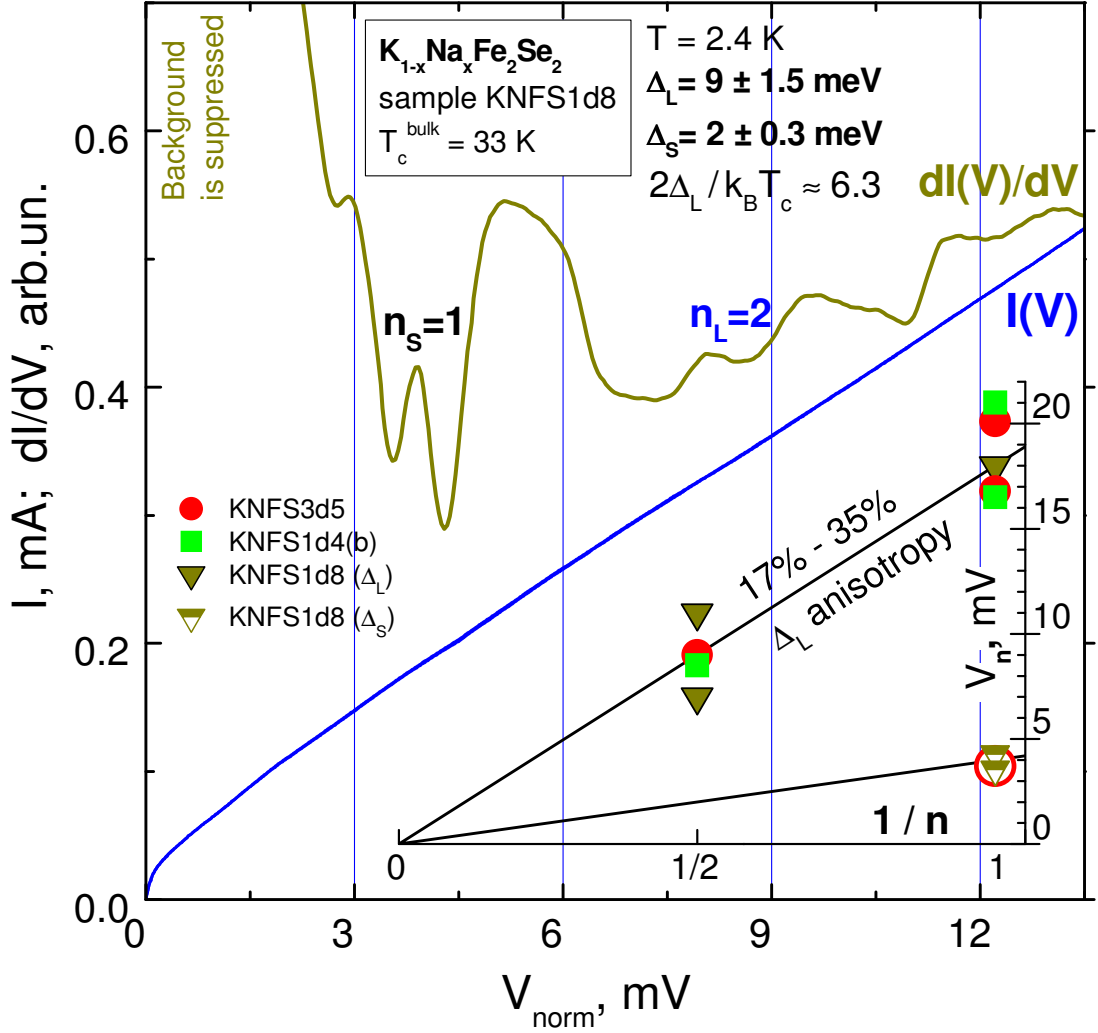

**Fig. 10**. I(V) (dark blue curve) and the dynamic conductance dI(V)/dV for SnS-contact KNFS1_d8 measured at T = 2.4 K. Exponential background in the dI(V)/dV-spectrum was suppressed. $n_L=2$ label marks the position of the second Andreev peculiarity for the large gap $\Delta_L^{aver} = (9 \pm 1.5)$ meV; $n_S=1$ label marks the first minimum corresponding to the small gap $\Delta_S = (2 \pm 0.3)$ meV. **The inset** shows the dependence between the Andreev minimum positions $V_n$ on their inversed subharmonic order $1/n$ for the large gap SGS (solid symbols; position for both minima in the doublets was taken) and the small gap (open symbols) in the spectra presented in Figs. 9, 10.

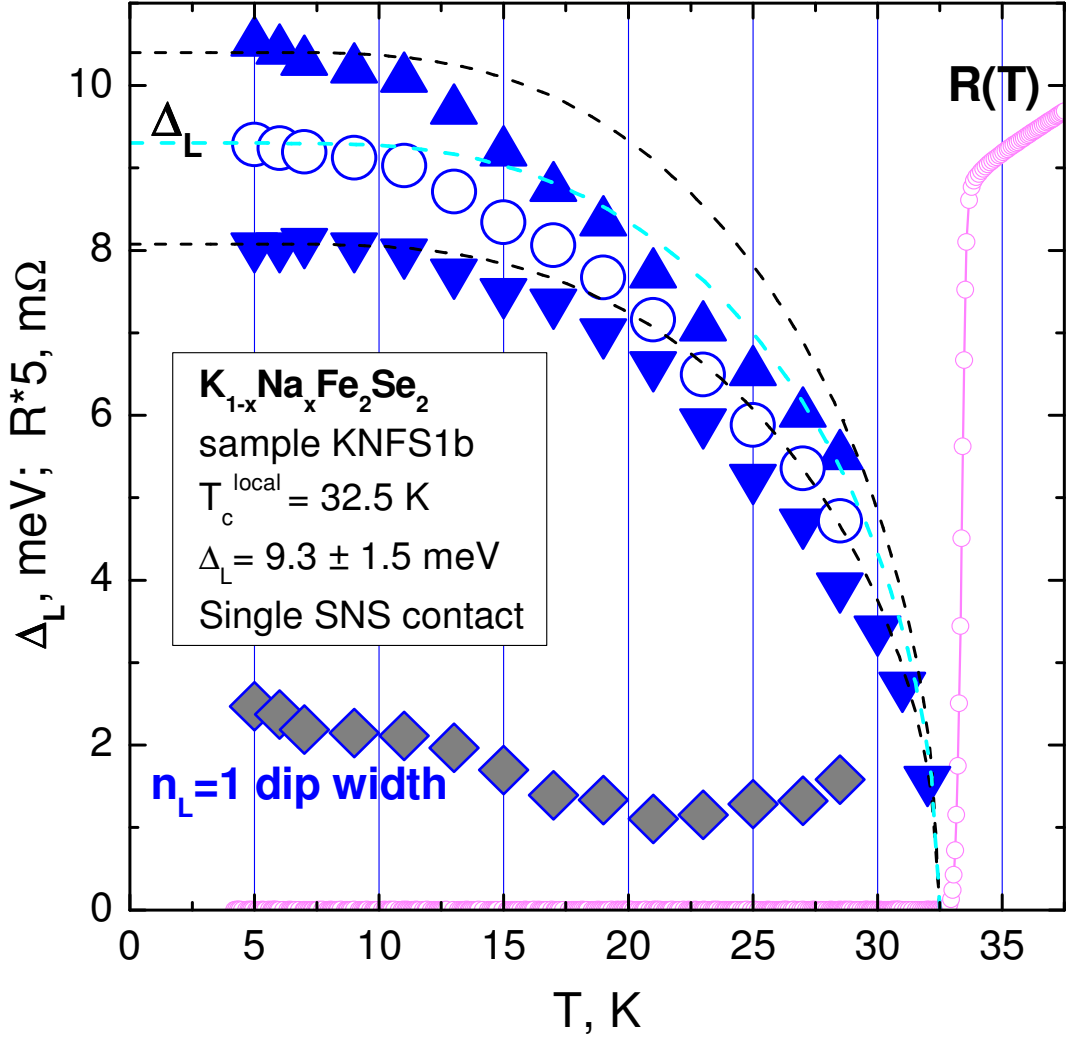

**Fig. 11**. Temperature dependence for the large gap doublet (up and down triangles) plotted on the base of KNFS1_d4 spectrum studies within $4.2\ K \leq T \leq T_c^{local} \approx 32.5\ K$. Open circles show the temperature behavior for the averaged value $\Delta_L^{aver}(4.2\ K) = (9.3 \pm 1.5)$ meV corresponding to the doublet center. Single-gap BCS-like curves (dashed lines) and R(T)-dependence (small circles) are presented for comparison. Gray rhombs depict the doublet width *vs*. T.